\documentclass[%
reprint,
superscriptaddress,
amsmath,amssymb,amsfonts,
aps,
prl,
]{revtex4-2}

\usepackage{graphicx}
\usepackage{dcolumn}
\usepackage{bm}
\usepackage{hyperref}
\usepackage[mathlines]{lineno}
\usepackage{sidecap}
\usepackage{verbatim}   
\usepackage[utf8x]{inputenc} 
\usepackage{xcolor}

\DeclareMathOperator\arctanh{arctanh}

\begin{document}

\title{Cluster-based Message-Passing (CluMP) Optimization for Complex QUBO Problems}

\author{Paolo Rissone}
    \email{rissone.paolo@gmail.com}
    \affiliation{Department of Physics, Sapienza University of Rome, P.le Aldo Moro 5, 00185, Rome (Italy)}

\author{Stefan Boettcher}
    \email{sboettc@emory.edu}
    \affiliation{Department of Physics, Emory University, 400 Dowman Drive, GA 30322, Atlanta (US)}

\author{Alfonso Amendola}
    \email{Alfonso.Amendola@eni.com}
    \affiliation{Eni S.p.A., Strada Statale 9 Via Emilia 1, San Donato Milanese I-20097 (Italy)}

\author{Simone Sala}
    \email{Simone.Sala@eni.com}
    \affiliation{Eni S.p.A., Strada Statale 9 Via Emilia 1, San Donato Milanese I-20097 (Italy)}

\author{Federico Ricci-Tersenghi}
    \email{federico.ricci@uniroma1.it}
    \affiliation{Department of Physics, Sapienza University of Rome, P.le Aldo Moro 5, 00185, Rome (Italy)}
    \affiliation{CNR-Nanotec, Rome unit, P.le Aldo Moro 5, 00185, Rome (Italy)}
    \affiliation{INFN-sezione di Roma, P.le Aldo Moro 5, 00185, Rome (Italy)}

\date{\today}

\begin{abstract}
Quadratic Unconstrained Boolean Optimization (QUBO) problems are widespread in both industrial applications and scientific studies. A QUBO problem corresponds to the optimization of a system of Ising spins defined on a generally sparse and heterogeneous graph. When the QUBO problem contains conflicting requests, the corresponding Ising system is frustrated, generating a complex energy landscape, which is hard to explore and optimize. Despite extensive algorithmic and hardware developments, finding low-energy configurations in these systems remains challenging (e.g., local-update heuristics typically become trapped in metastable states), especially when the (possibly frustrated) interactions generate extended correlated domains.

We introduce CluMP (Cluster-based Message-Passing), an algorithm that performs collective updates on connected clusters of spins using information from Belief Propagation (BP). By controlling the amount of frustration within clusters, CluMP enables BP convergence on large subgraphs and proposes nonlocal rearrangements involving up to hundreds of spins in a single move. We benchmark CluMP against state-of-the-art local-update heuristics on spin-glass models defined on several graph topologies, including random regular graphs and lattice regular graphs in two and three dimensions. Cluster moves consistently bypass local trapping and reach lower energies with fewer effective operations than single-spin dynamics. These results demonstrate that frustration-tolerant cluster updates can be implemented efficiently on sparse graphs. 
The CluMP framework provides a scalable strategy for large-scale combinatorial optimization and inference problems, where exploiting medium- and long-range correlations is key to navigating complex energy landscapes.
\end{abstract}

\maketitle

\paragraph{Introduction.}
Optimization in rugged high-dimensional landscapes is an ubiquitous challenge across physics, computer science, and real-world applications \cite{lucas2014ising}. Problems such as Max-Cut, community detection, and spin-glass ground state (GS) search require navigating configuration spaces structured by metastable basins separated by large barriers. When formulated as interacting variables on complex graphs, frustration and disorder generate a proliferation of local minima that severely hinders exploration by standard stochastic algorithms \cite{barahona1982computational}.

Markov-Chain Monte Carlo methods typically rely on single-variable updates to satisfy the detailed balance condition \cite{metropolis1953equation,hastings1970monte}. While ensuring convergence to the right asymptotic measure (e.g., the Gibbs measure in statistical physics), the microscopic reversibility constrains the dynamics: correlations propagate diffusively, leading to critical slowing down and exponentially suppressed barrier crossing in glassy systems. Heuristic methods such as Simulated Annealing (SA) \cite{kirkpatrick1983optimization} and Extremal Optimization (EO) \cite{boettcher2000extremal,boettcher2004new} accelerate convergence to low-energy states by skipping the requirement of sampling the asymptotic measure, e.g., by working in the out-of-equilibrium regime and/or by breaking detailed balance. These heuristics show strong empirical performance on large-scale benchmarks \cite{angelini2019monte,caracciolo2023simulated,angelini2023limits,angelini2026algorithmic}. Yet, they remain fundamentally local: each step modifies a single spin, limiting the reorganization of extended correlated domains.

Collective Algorithms (CAs) address this limitation by performing nonlocal updates. Cluster Monte Carlo schemes, such as the Swendsen-Wang and Wolff algorithms \cite{swendsen1987nonuniversal,wolff1989collective,wang1990cluster}, exploit correlation structures to generate extended moves and eliminate critical slowdown in unfrustrated systems through the Fortuin-Kasteleyn-Coniglio-Klein (FKCK) construction \cite{fortuin1972random,coniglio1980clusters}. However, frustration breaks this mechanism: competing couplings invalidate the FKCK construction, leading to pathological percolation and loss of dynamical speedup \cite{kandel1991general,cataudella1993cluster}. Replica-based cluster constructions improve percolation properties \cite{munster2023cluster}, but so far remain effective only in two-dimensional lattices \cite{houdayer2001cluster,houdayer2004low} or in restricted geometries combined with parallel tempering \cite{bernaschi2026low}.

Therefore, a scalable CA for frustrated systems is still missing. The cluster-exact approximation (CEA) \cite{hartmann1996cluster} partially addresses the problem by constructing frustration-free clusters whose GS can be computed exactly, while freezing the remainder of the system. Extensions based on Belief Propagation (BP) \cite{decelle2014belief} enable finite-temperature sampling, but remain confined to frustration-free clusters. Introducing frustration typically causes BP to lose convergence \cite{lage2011inference}, consistent with the fact that BP is exact only on trees and uncontrolled on loopy, frustrated graphs. Quantum-guided cluster proposals \cite{finvzgar2024quantum,eder2025quantum} pursue collective updates for combinatorial optimization, yet current implementations are restricted to small instances, several orders of magnitude below the spin-glass and random-graph benchmarks used in the present work.

Here, we introduce Cluster-based Message-Passing (CluMP), a heuristic algorithm that performs collective updates on connected clusters with an adjustable degree of frustration, guided by BP marginals. In contrast to previous approaches, CluMP tolerates substantial frustration and generates clusters of hundreds of spins. The method combines large-scale nonlocal rearrangements with the robustness of BP on near-tree structures, enabling efficient access to low-energy configurations that remain inaccessible to purely local dynamics. By demonstrating scalable, frustration-tolerant cluster moves, CluMP closes a longstanding gap between collective algorithms and practical large-scale optimization of complex systems.


%
\begin{figure}[!t]
    \centering
    \includegraphics[width=\columnwidth]{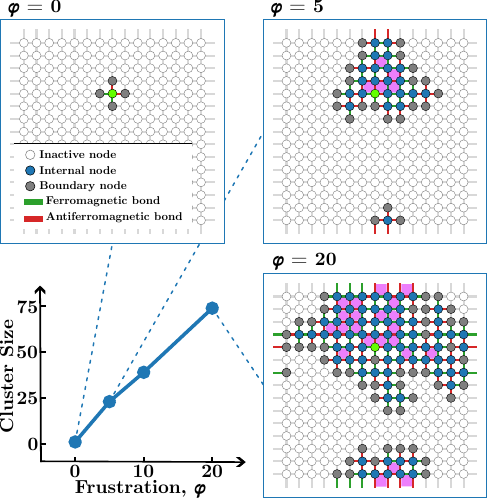}
    \caption{\textbf{Clusters at controlled frustration.} Starting from a random seed (green), a connected cluster is grown until the number of frustrated loops reaches the threshold, $\phi$. A loop is frustrated when the product of its couplings is negative. Examples on a 2d LRG show that increasing $\phi$ yields larger clusters (blue nodes) and a higher number of frustrated loops (violet plaquettes). Boundary nodes (grey), adjacent to at least one external edge but not belonging to the cluster, define the interface with the rest of the graph. Note that for $\phi = 0$, the cluster reduces to a single node.}
    \label{fig:FIG1}
\end{figure}

\paragraph{The CluMP Algorithm.} Consider a spin system on a graph $\mathcal{G}(V, E)$ with nodes $V$ and edges $E$, where each node carries an Ising spin $\sigma_i=\pm 1$ and edges host couplings, $J_{ij}$, which may be randomly drawn from a given probability distribution. The Hamiltonian is
\begin{equation}
    \mathcal{H}(\boldsymbol{\sigma})=-\sum_{\langle ij\rangle} J_{ij}\,\sigma_i\sigma_j,
    \label{eq:hamiltonian}
\end{equation}
typical of spin–glass models \cite{castellani2005spin, parisi2006spin}.  
For such a graph, a cluster $\mathcal{C}$ is grown from a random seed and iteratively expanded by adding neighboring nodes until the number of elementary frustrated loops, i.e., minimal cycles (square plaquettes in $d=2$) with a negative couplings product, reaches a maximum threshold $\phi$. Once the growth stops, boundary nodes are defined as the vertices not belonging to the cluster $\mathcal C$ that are adjacent to at least one node in $\mathcal C$.
Figure~\ref{fig:FIG1} shows the cluster growth on a $d=2$ lattice with increasing $\phi$, where blue (gray) vertices are internal (boundary) nodes.
Given a cluster $\mathcal{C}$, our aim is to find the GS in $\mathcal{C}$, given the fixed boundary nodes. This is achieved by running the BP algorithm at zero temperature (see End Matter \cite{EndMatter}) on the internal nodes. Boundary nodes act as an effective external field: they send messages into the cluster but do not receive any in return. We emphasize that the initialization of both cluster and boundary nodes is crucial for BP stability (see End Matter \cite{EndMatter}).

\begin{figure*}[t]
    \centering
    \includegraphics[width=\textwidth]{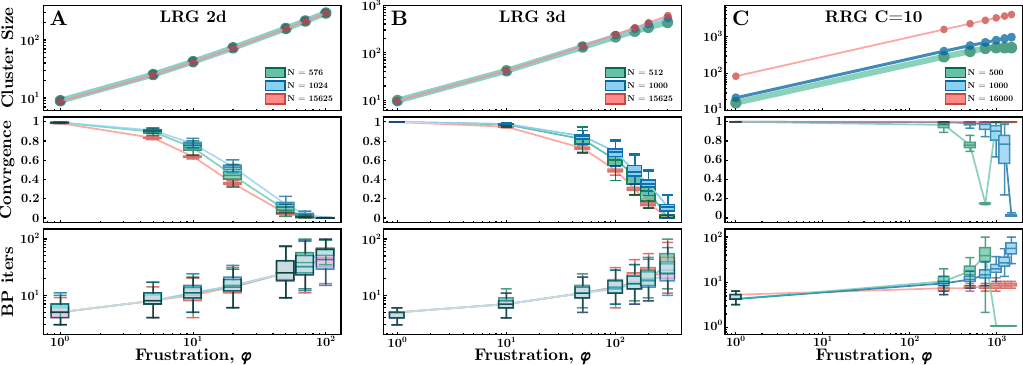}
    \caption{\textbf{BP on frustrated clusters.} Zero-temperature BP updates (Eqs. \eqref{eq:BPzeroT}) on clusters with increasing frustration $\phi$ for 2d and 3d lattice regular graphs (LRGs; \textbf{A,B}) and random regular graphs with $C=10$ (RRGs; \textbf{C}), at $V\sim10^2$–$10^4$ (100 runs, $2^8$ iterations). LRGs show rapid loss of convergence with increasing $\phi$ (2d most sensitive), while RRGs remain stable up to $\phi\sim10^3$, allowing clusters up to $|\mathcal{C}|\sim0.8V$. Box plots show medians, interquartile ranges (IQRs), and 1.5 IQR whiskers.}
    \label{fig:FIG2}
\end{figure*}

In Fig.~\ref{fig:FIG2}, we show the BP performance on square ($d=2$) and cubic ($d=3$) lattice regular graphs (LRGs) and on random regular graphs (RRGs) with $C=10$. For each graph type, we performed $100$ independent runs of $2^8$ iterations while varying the frustration threshold $\phi$ and the system size $V \sim 10^2-10^4$. On LRGs (panels A and B), the cluster size $|\mathcal{C}|$ grows with $\phi$ with a negligible dependence on $V$. Two-dimensional lattices are the most fragile: BP convergence drops abruptly for $\phi\gtrsim 20$, when clusters reach $\mathcal{O}(10^2)$ nodes. In contrast, for RRGs (panel C), $|\mathcal{C}|$ increases with $V$ and BP converges on clusters of size up to $|\mathcal{C}|\!\sim\!0.8V$, with only a weak dependence on $\phi$.
These trends follow directly from the structure of frustrated loops. LRGs contain a high density of short loops---about half of which are frustrated---quickly exhausting the frustration budget and destabilizing BP fixed points. RRGs, instead, have loop lengths growing as $\mathcal{O}(\log V)$ and a vanishing density of short cycles \cite{lucibello2014finite}, resulting in far fewer frustrated substructures and allowing BP to converge on large clusters. When BP fails on RRGs (Fig.~\ref{fig:FIG2}C, middle), the cluster has grown to almost the full graph, a regime in which zero–temperature BP is known not to converge for spin glass models.

Once BP has reached convergence in $t = t^*$ iterations, the CluMP algorithm assigns a new spin configuration to cluster nodes according to their marginals
\begin{equation}
    \sigma_i = \mathrm{sign}\!\left( \sum_{k \in \partial i} u^{(t^*)}_{k \rightarrow i} \right),
    \label{eq:set_spins}
\end{equation}
where the sum runs over all BP messages $\{u^{t^*}_{k \rightarrow i}\}$ arriving in $i$ from its neighbors $k\in\partial i$.
Cluster generation and subsequent BP-based spin updates form the core of the algorithm. Repeating this procedure for multiple iterations creates overlapping clusters that efficiently explore the configuration space under bounded frustration.

We have also introduced an algorithm variant inspired by the population annealing method \cite{machta2010population}, denoted Resampling-CluMP (R-CluMP), where a population of $R$ replicas evolves in parallel. At each iteration, an independent cluster is grown on every replica and its spins are updated via the BP rule in Eq. \eqref{eq:set_spins}. Every $\Delta n = 100$ iterations, the population is resampled to bias the ensemble toward low-energy states. Each replica is assigned a weight $w_r = \exp\!\left[-\frac{E_r - E_{\min}}{\sigma_E}\right]$, where $E_r$ is the energy of replica $r$, $E_{\min}$ the lowest energy in the population, and $\sigma_E$ the standard deviation of the population energy. Systematic resampling is then performed by generating $R$ equally spaced targets on the cumulative distribution $W_r = \sum_{k \le r} w_k$, duplicating high-weight (low-energy) replicas and eliminating low-weight ones.
To preserve diversity, replicated configurations undergo a local mutation step after resampling \cite{liang2000evolutionary}: spins in the cluster are flipped with a small probability $\mathcal{O}(10^{-2})$. The population then continues to evolve through cluster growth and BP updates until the next resampling stage.

In what follows, we benchmark CluMP and R-CluMP on LRGs and RRGs of size $V\sim10^3$ and Gaussian-distributed couplings. The frustration threshold $\phi$ is set to yield $\sim50\%$ BP convergence (cutoff at $10^2$ iterations), corresponding to a typical convergence times of $10$–$50$ iterations (Fig.~\ref{fig:FIG2}, bottom): $\phi\approx 20$ for $d=2$ LRGs, $\phi\approx 120$ for $d=3$ LRGs, and $\phi\approx 1250$ for RRGs \footnote{For other values of $V$ the optimal $\phi$ is unchanged for LRGs and should be scaled by $V$ for RRGs.}. CluMP runs for $2^{14}$ iterations, while R-CluMP uses $2^{12}$ iterations with $R=10$ replicas.
We compare against single-spin heuristics: SA and $\tau$EO \cite{boettcher2000nature,boettcher2002optimization}, which is widely considered the state-of-the-art approach for spin-glass optimization.
For $\tau$EO, we set the control parameter $\tau=1.5$, 2.5 and 1.25 for 2d, 3d and RRG, respectively. For SA, after several preliminary tests, we have opted for an effective linear schedule in the inverse temperature $\beta$, where $n^{\rm SA}$ Monte Carlo sweeps are performed between $\beta_\text{init}=0$ and $\beta_\text{final}$ (one sweep per temperature); the optimal parameters used are $(\beta_\text{final},n^{\rm SA})=(10,2^{18}),(10,2^{16}),(5,2^{18})$ for 2d, 3d and RRG, respectively.
All these details are reported in the GitHub repository \cite{CluMP2026code}.
Statistics are averaged over 10 samples, with 100 runs per sample for CluMP, R-CluMP, and SA, and 10 runs for $\tau$EO due to longer convergence times.

%
\begin{figure*}[t]
    \centering
    \includegraphics[width=\textwidth]{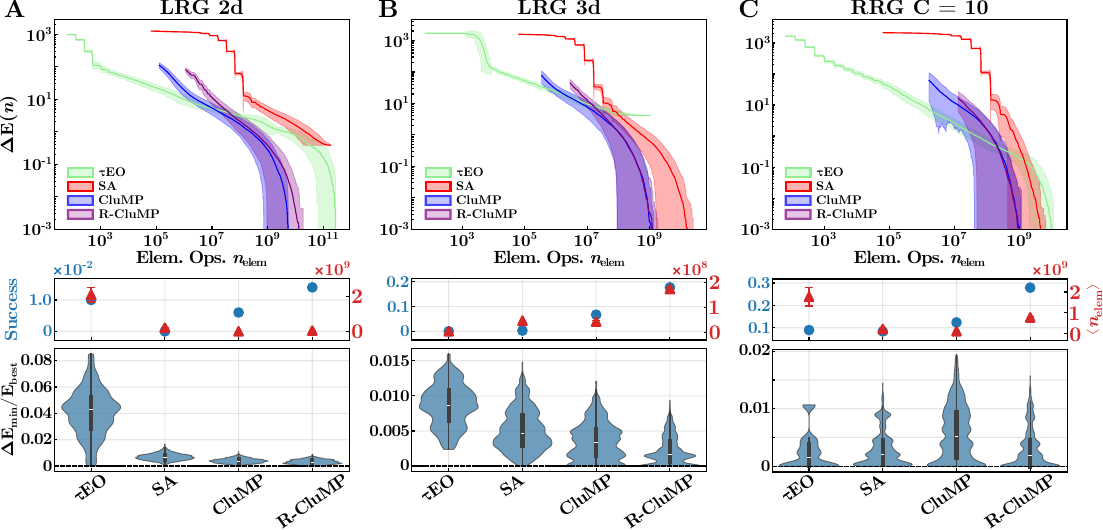}
    \caption{\textbf{CluMP performance.} Performance of $\tau$EO, SA, CluMP, and R-CluMP on LRGs ($d=2,3$; \textbf{A,B}) and RRGs (\textbf{C}). Ten samples per graph family, $10^2$ runs per sample. Top: resampled convergence trajectories show the best-so-far energy as a function of the number of elementary operations, $n_\text{elem}$ (see text). Middle: success rate (blue circles) and average number of effective single-spin iterations (red triangles) to reach $E_{\rm best}$. Bottom: violin plots of the normalized gap $\Delta E_{\min}/E_{\rm best}$ with embedded box plots (median, IQR, 1.5 IQR whiskers). Cluster methods achieve smaller gaps and higher success rates, with the largest speedup on LRGs.}
    \label{fig:FIG3}
\end{figure*}

\paragraph{Results.} Figure~\ref{fig:FIG3} compares all algorithms on LRGs ($d=2,3$) and RRGs with $C=10$. To enable a direct comparison between local and cluster-based dynamics, performance is measured by the number of elementary operations, $n_{\rm elem}$, which quantifies the effective computational time. For CluMP and R-CluMP, $n^{\rm CluMP}_{\rm elem} = n \, \langle |\mathcal{C}| \rangle \, \langle n_{\rm BP} \rangle \, R$, where $n$ is the number of cluster updates, $\langle |\mathcal{C}| \rangle$ the mean cluster size, $\langle n_{\rm BP} \rangle$ the average BP iterations per update, and $R$ the number of replicas. For SA, $n^{\rm SA}_{\rm elem}=n\,V$. $\tau$EO is expressed directly in elementary operations.
To assess the typical rate of convergence (top panels), we proceed as follows for each algorithm: we concatenate the outputs of all the runs, and we compute the best-so-far energy as a function of $n_\text{elem}$; we repeat the procedure over $10^3$ random permutations of the runs order, to obtain the mean decay rate and uncertainty bounds. This resampling produces a reliable comparison of optimization efficiency across algorithms by quantifying the computational effort required to approach the GS in a very robust way.
An aware reader may notice that Population Annealing (PA) is not included in the comparison. We have tested PA as well, but to become competitive, PA requires $R=\mathcal{O}(10^2$–$10^4)$ replicas, implying $n_{\rm elem}$ orders of magnitude larger than the methods shown here (see Supp. Material \cite{SuppMat}). 
\begin{table}[t]
\centering
\begin{tabular}{lccc}
\hline
 & Energy Gap ($\cdot 10^{-2}$) & Elem. Ops. ($\cdot 10^8$) & Success\\
\hline
$\tau$EO & $1.7\,(8)$ & $13\,(3)$ & 0.03\\
SA       & $0.5\,(5)$ & $1.64\,(4)$ & 0.03\\
CluMP    & $0.4\,(6)$ & \textbf{0.62\,(4)} & 0.07\\
R-CluMP  & \textbf{0.3(4)} & $3.4\,(2)$ & \textbf{0.16}\\
\hline
\end{tabular}
\caption{\textbf{CluMP global performance.} Performance of the different algorithms averaged over all graph families, samples, and runs. Reported are the mean normalized gap $\langle \Delta E_{\min}/E_{\rm best} \rangle$ (col.~1), the average number of elementary operations $\langle n_{\rm elem}\rangle$ to reach $E_{\rm best}$ (col.~2), and the success rate (col.~3), summarizing the statistics of Fig.~\ref{fig:FIG3} (bottom panels).}
\label{TAB:table1}
\end{table}

Cluster updates consistently outperform local-move algorithms, yielding a speedup of $\mathcal{O}(10^2)$ on $d=2$ LRGs (Fig.~\ref{fig:FIG3}A) and about a factor of two on $d=3$ LRGs and RRGs (Fig.~\ref{fig:FIG3}B,C).
The overall performance (bottom panels) is summarized by violin plots of the normalized energy gap $\Delta E_{\min} / E_{\rm best} = (E_{\min} - E_{\rm best}) / E_{\rm best}$, where $E_{\min}$ is the lowest energy reached by a given run of a given algorithm, and $E_{\rm best}$ is the lowest energy found across all algorithms and runs. Success rates and average number of elementary operations, $\langle n_{\rm elem} \rangle$, (middle panels) are obtained by counting runs with $\Delta E_{\min}=0$ and averaging the time needed to reach $E_{\rm best}$.
The results shown in Fig.~\ref{fig:FIG3} are for graphs of increasing connectivity ($C=4,6,10$), which represent more challenging instances. For lower-connectivity random graphs (e.g., $C=4$ RRGs), performance improves substantially (see Supp. Material \cite{SuppMat}), with speedups reaching $\mathcal{O}(10^4)$ and a 40\% success rate.

As summarized in Table~\ref{TAB:table1}, CluMP and R-CluMP achieve the smallest $\Delta E_{\min}$, the highest success rates, and the lowest $\langle n_{\rm elem} \rangle$, thus representing a clear success for cluster-based optimization in sparse and heterogeneous QUBO problems. It is fair to say that this advantage depends on connectivity: as graphs become denser, single-spin methods progressively narrow the gap (see Supp. Material \cite{SuppMat}), but for very sparse problems, we believe CluMP and R-CluMP set the new standard to compare with.

\begin{figure*}[t]
    \centering
    \includegraphics[width=\textwidth]{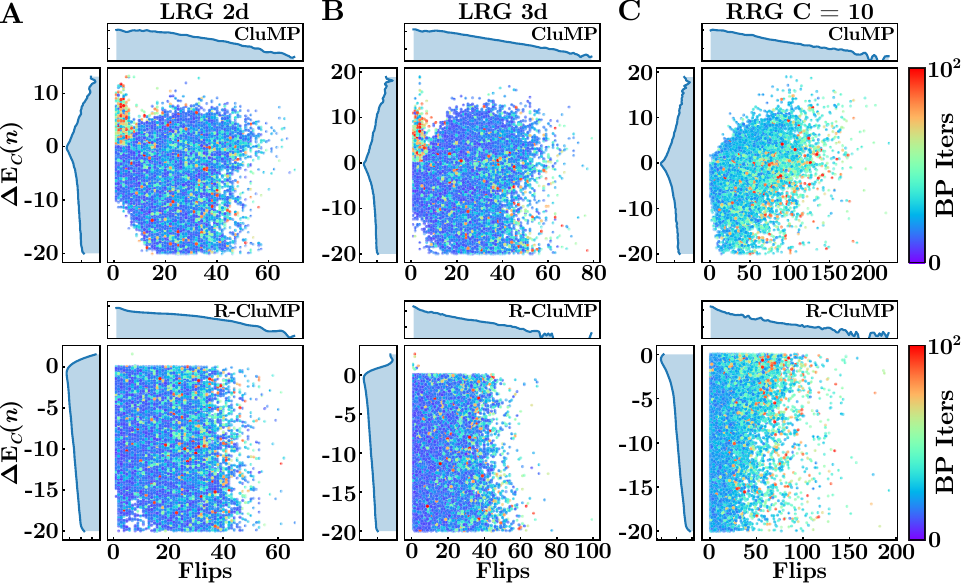}
    \caption{\textbf{Statistics of CluMP updates.} Correlation between cluster energy change $\Delta E_{\mathcal{C}}(n)$ and number of flipped spins for LRGs in 2d (\textbf{A}), in 3d (\textbf{B}), and $C=10$ RRGs (\textbf{C}). Data aggregate all iterations, runs, and samples for CluMP (top) and R-CluMP (bottom). Marginals show distributions of $\Delta E_{\mathcal{C}}$ and flip counts; colors indicate BP iterations to convergence (blue: fast, red: slow). Cluster moves may flip up to $\mathcal{O}(10^2)$ spins, largely independent of $\Delta E_{\mathcal{C}}$, enabling single-cluster-flip transitions between nearly isoenergetic states. In LRGs, slow BP convergence correlates with $\Delta E_{\mathcal{C}}>0$; in RRGs, this correlation weakens. Resampling in R-CluMP strongly suppresses $\Delta E_{\mathcal{C}}>0$ updates.}
    \label{fig:FIG4}
\end{figure*}

The statistical analysis of single cluster updates in CluMP and R-CluMP is presented in Fig.~\ref{fig:FIG4}, where the scattered plots show the correlation between the number of spins flipped in a single cluster update and the corresponding change in the extensive energy. It is worth stressing that a single cluster update may flip $\mathcal{O}(10^2)$ spins, largely independent of the associated energy change $\Delta E_{\mathcal{C}}(n)$. Even more importantly, these cluster updates produce large rearrangements that often connect nearly isoenergetic configurations ($\Delta E_{\mathcal{C}} \approx 0$), effectively bypassing the barriers that confine single-spin dynamics.

The convergence time of BP (color scale in Fig.~\ref{fig:FIG4}) reveals structural differences between ensembles. On LRGs, slow convergence correlates with $\Delta E_{\mathcal{C}}>0$ and small flip counts, signaling difficulties when short frustrated loops are many. On RRGs, this correlation weakens: slow convergence mainly reflects the intrinsic cost of optimizing larger, more frustrated clusters. Resampling in R-CluMP suppresses energetically unfavorable updates, biasing the dynamics toward downhill or isoenergetic transitions while preserving the exploratory power of large moves and stabilizing the low-energy population.


\paragraph{Discussion and Conclusions.}
In this work, we have introduced a class of optimization algorithms that can efficiently update large clusters of spins altogether, even in strongly disordered models, like spin glasses, thus outperforming other optimization algorithms in the search for the solution to complex QUBO problems.
This achievement shows that efficient cluster algorithms for spin glasses can exist in general and not only restricted to some specific topologies, as previously believed for the so-called Houdayer cluster move \cite{houdayer2001cluster, bernaschi2026low} \footnote{While completing this work, a new proposal to use the Houdayer cluster move has been shown effective in $d=3$ \cite{chilin2026cluster}}.

The CluMP algorithm can grow clusters of a variable size, controlled by the amount of internal frustration, and then compute the GS configuration in the cluster via the BP algorithm run at zero temperature. The great advantage of BP is that it can work even in the presence of a limited amount of frustration and short loops \cite{lage2011inference, dominguez2011characterizing, lage2014message}, thus providing single-spin marginals, which correspond to the GS configuration.
Previous attempts to compute a local GS were limited to unfrustrated regions \cite{hartmann1996cluster}, but this is not enough to optimize a strongly disordered model.

Our analysis shows that cluster size grows predictably with the frustration level $\phi$. On LRGs, $|\mathcal{C}|$ is essentially independent of the system size, and can grow to several tens of spins before BP convergence rate starts degrading. On RRGs, the maximum cluster size grows linearly with $V$, reflecting the scarcity of short loops and the resulting locally tree-like structure (Fig.~\ref{fig:FIG2}).

These collective moves provide a clear optimization advantage against local-updates methods: both CluMP variants reach lower energies with substantially fewer iterations (Fig.~\ref{fig:FIG3}), underscoring their ability to navigate rugged, highly frustrated landscapes. Remarkably, individual cluster updates often induce near-isoenergetic rearrangements ($\Delta E \ll 1$, see Fig.~\ref{fig:FIG4}), demonstrating the capability of collective moves to cross shallow barriers that impede local optimization schemes. 
Moreover, the R-CluMP variant, by systematically resampling a population of replicas, strongly reduces the possibility of cluster updates with $\Delta E > 0$ (lower panels in Fig.~\ref{fig:FIG4}).

Taken together, our results show that collective-move optimization can be extended to generic spin glass models and complex QUBO problems.
However, the graph properties may influence the advantage of cluster updates.
For example, we have noticed that integer couplings and higher connectivities may worsen the performance of CluMP (see Supp. Material \cite{SuppMat}).
Further studies are probably needed to improve the cluster generation in these situation, with the aim of achieving a fully general purpose cluster update algorithm for strongly disordered models.

In this first study, we have focused on the optimization problem of finding the GS. Future developments will be dedicated to satisfy the detailed balance condition, thus converting CluMP into an efficient sampler.


\paragraph{Acknowledgments.} The research has received financial support from the “National Center for HPC, Big Data and Quantum Computing”, Project CN\_00000013, CUP B83C22002940006, NRRP Mission 4 Component 2 Investment 1.4,  Funded by the European Union - NextGenerationEU. P.~Rissone has also been supported by the Sapienza University starting grant “Avvio alla Ricerca 2025”, CUP B83C25004300005.

\paragraph{Data Availability.} The code used in this study is publicly available at \cite{CluMP2026code}. The datasets generated and analyzed during this study are publicly available at \cite{CluMP2026data}.


\bibliography{Bibliography}


\onecolumngrid

\section*{End Matter}

\subsection*{Belief Propagation: Update Equations at Zero Temperature}

Belief Propagation (BP) is an iterative message-passing algorithm that aims at solving the self-consistency cavity equations \cite{mezard2003cavity} and computing local marginals (i.e., magnetizations and nearest neighbors correlations).
The BP algorithm works by exchanging messages between neighboring variables, until a fixed point is eventually reached. BP is a deterministic algorithm, whose fate depends solely on the interaction graph and the initial condition: BP may (i) not converge or (ii) reach a unique or (iii) multiple fixed points \cite{perugini2018improved}.

Let us consider a Hamiltonian (energy) function with Ising spins $\sigma_i = \pm 1$ and pairwise interactions defined on a generic graph $G=(V,E)$
\begin{equation}
    H(\boldsymbol{\sigma}) = - \sum_{(ij)\in E} J_{ij} \sigma_i \sigma_j + \sum_{i\in V} H_i \sigma_i \, ,
\end{equation}
where the external field $H_i$ and coupling $J_{ij}$ are given (even if eventually extracted from a generic probability distribution).
At a finite temperature $T$, the BP equations to updates the messages are the following
\begin{equation}  
    \label{eq:BPfiniteT}
    \begin{split}
        h^{(t+1)}_{i \rightarrow j} &= H_i + \sum_{k \in \partial i \setminus j} u^{(t)}_{k \rightarrow i}\,, \\
        u^{(t)}_{k \rightarrow i} &= T \, \arctanh \Big[ \tanh(\beta J_{ki}) \tanh(\beta h^{(t)}_{k \rightarrow i}) \Big] \, ,
    \end{split}
\end{equation}
where $\beta = 1/T$ and $\partial i$ is the set of neighbors of spin $i$.
At convergence, the single spin marginals are given by
\begin{equation}  
        m_i = \langle s_i \rangle = \tanh\bigg[\beta \Big(H_i + \sum_{k \in \partial i} u^*_{k \rightarrow i}\Big)\bigg]\,.
\end{equation}
In the zero temperature limit, the above equations further simplify to
\begin{equation}  
    \label{eq:BPzeroT}
    \begin{split}
        h^{(t+1)}_{i \rightarrow j} &= H_i + \sum_{k \in \partial i \setminus j} u^{(t)}_{k \rightarrow i}\,, \\
        u^{(t)}_{k \rightarrow i} &= \hat u_{J_{ki}} \big( h^{(t)}_{k \rightarrow i} \big)\,,
    \end{split}
\end{equation}
with
\begin{equation}  
    \label{eq:u_J}
    \hat u_J(x) = 
    \begin{cases}
        -J & x \le -J\,,\\
        x & -J < x < J\,,\\
        J & x \ge J\,.
    \end{cases}
\end{equation}
These zero-temperature update rules for the BP messages are used in the CluMP algorithm to propagate the effective fields in the cluster and eventually propose---in case of BP convergence---the collective spin updates according to
\begin{equation}  
        m_i = \text{sign}\bigg[H_i + \sum_{k \in \partial i} u^*_{k \rightarrow i}\bigg]\,.
\end{equation}

\subsection*{Boundary Conditions and BP Convergence}

The convergence of zero-temperature BP within a given cluster is highly sensitive to the boundary conditions imposed by the external spins. For continuous couplings, such as Gaussian-distributed $J_{ij}$, convergence is significantly enhanced by applying strong boundary fields aligned with the external spin configuration (hard boundary conditions). These large effective fields stabilize the message dynamics and promote the selection of a consistent fixed point within the cluster.

\begin{figure*}[h]
    \centering
    \includegraphics[width=0.8\textwidth]{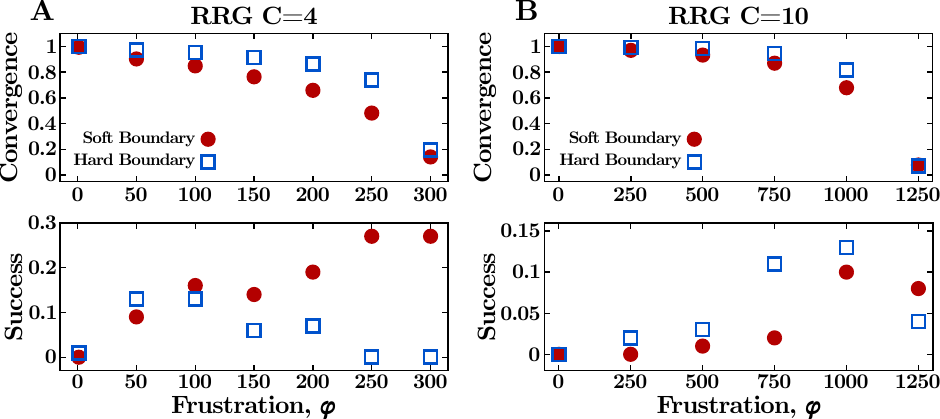}
    \caption{\textbf{CluMP performance under different boundary conditions.} Effects of hard (blue squares) and soft (red circles) BP boundary conditions in RRGs with connectivity $C=4$ (\textbf{A}) and $C=10$ (\textbf{B}). Hard boundary conditions generally yield higher BP convergence rates (top panels), but significantly degrade CluMP performance at low connectivity (bottom panels), effectively reducing the maximum frustration threshold and limiting the accessible cluster size. In contrast, soft boundary conditions produce more robust and reproducible performance across different graph instances.} 
    \label{fig:FIGEM1}
\end{figure*}

In contrast, for discrete couplings ($J_{ij}=\pm1$), hard boundary conditions lead to strongly graph-dependent behavior. Although they yield high BP convergence rates (Fig.~\ref{fig:FIGEM1}, top), they result in a reduced success probability (fraction of runs reaching the true ground state) particularly at low connectivity (Fig.~\ref{fig:FIGEM1}, bottom).
In this case, better performance is obtained by introducing soft stochastic boundary fields, $h_{k\to i} = \sigma_k [1 - C_s \cdot\mathrm{rand}(0,1)]$, where $\sigma_k$ is the external boundary spin and $C_s=\mathcal{O}(10^{-4})$. These controlled fluctuations perturb local degeneracies at the cluster boundary, enabling the BP dynamics to escape unfavorable attractors and converge toward a fixed-point solution. This substantially reduces graph-to-graph fluctuations and improves the overall performance of the algorithm.

This marked sensitivity to boundary conditions arises from the deterministic character of zero-temperature BP updates (Eq. \eqref{eq:u_J}]) and from the presence of multiple competing fixed points in frustrated subgraphs \cite{perugini2018improved}. Degeneracies are particularly abundant for integer couplings, where rigid boundary fields can enforce incompatible constraints and induce oscillatory or non-convergent message dynamics. 

For this reason, we focus on Gaussian-distributed couplings, for which hard boundary conditions provide stable and reproducible convergence within frustration-controlled clusters. Results for RRGs with integer couplings ($C=4,10$) are reported in the Supp. Material \cite{SuppMat}.

\end{document}


\title{Supplemental Material:\\ Cluster-based Message-Passing (CluMP) Optimization for Complex QUBO Problems}

\author{Paolo Rissone}
    \email{rissone.paolo@gmail.com}
    \affiliation{Department of Physics, Sapienza University of Rome, P.le Aldo Moro 5, 00185, Rome (Italy)}

\author{Stefan Boettcher}
    \email{sboettc@emory.edu}
    \affiliation{Department of Physics, Emory University, 400 Dowman Drive, GA 30322, Atlanta (US)}

\author{Alfonso Amendola}%
    \email{Alfonso.Amendola@eni.com}
    \affiliation{Eni S.p.A., Strada Statale 9 Via Emilia 1, San Donato Milanese I-20097 (Italy)}

\author{Simone Sala}%
    \email{Simone.Sala@eni.com}
    \affiliation{Eni S.p.A., Strada Statale 9 Via Emilia 1, San Donato Milanese I-20097 (Italy)}

\author{Federico Ricci-Tersenghi}%
    \email{federico.ricci@uniroma1.it}
    \affiliation{Department of Physics, Sapienza University of Rome, P.le Aldo Moro 5, 00185, Rome (Italy)}
    \affiliation{CNR-Nanotec, Rome unit, P.le Aldo Moro 5, 00185, Rome (Italy)}
    \affiliation{INFN-Rome 1 section, P.le Aldo Moro 5, 00185, Rome (Italy)}

\date{\today}

\maketitle
\onecolumngrid
\tableofcontents

\section{APPENDIX A: Comparison with Population Annealing}

Population Annealing (PA) \cite{machta2010population} extends standard Simulated Annealing (SA) by evolving a population of $R$ replicas through a common annealing schedule. At every temperature step, each replica undergoes the same local Metropolis updates used in SA, followed by a resampling stage in which replicas are selected according to their Boltzmann weights. In the limit of a large population size $R$, this procedure suppresses trapping in metastable states and improves equilibration, potentially allowing the algorithm to reach the true ground state.

To benchmark its performance, we compared PA with SA and CluMP on lattice regular graphs (LRGs) in $d=2$ and $d=3$, using the same disorder ensembles and optimization protocol adopted in the main text. For a direct comparison, PA was implemented using the same temperature schedule, number of temperature steps, and local Monte Carlo sweeps as SA, with the optimal annealing parameters $(\beta_{\rm final},n^{\rm SA})=(10,2^{18})$ and $(10,2^{16})$ in $d=2$ and $d=3$, respectively. We considered population sizes ranging from $R=10$ to $R=1000$.

\begin{figure*}[t]
    \centering
    \includegraphics[width=0.9\textwidth]{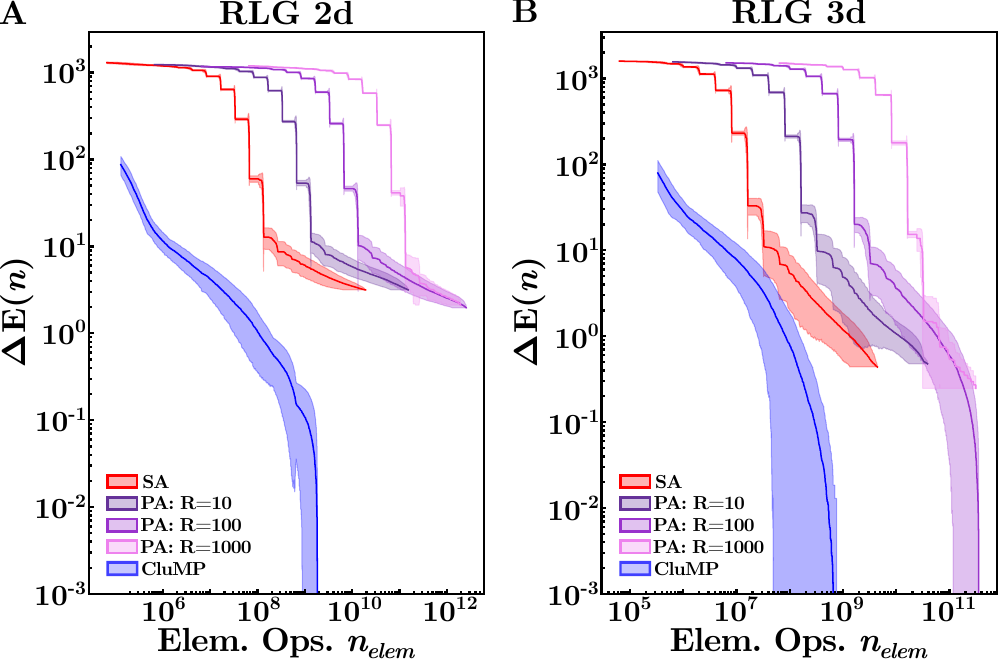}
    \caption{\textbf{Population Annealing benchmark on lattice regular graphs.} Performance of Population Annealing (PA), compared with Simulated Annealing (SA) and CluMP, on lattice regular graphs (LRGs) in $d=2$ (\textbf{A}) and $d=3$ (\textbf{B}). PA is implemented using the same annealing schedule and local Monte Carlo dynamics as SA (see text), with population sizes $R=10,100,1000$. Increasing $R$ systematically improves the optimization performance of PA, progressively reducing the residual energy gap $\Delta E$ (look at the heights of the steps) and reaching $\Delta E=0$ in the three-dimensional case. However, the linear scaling of the computational cost with $R$, makes the population sizes required to match CluMP computationally expensive, leading to effective runtimes orders of magnitude larger.}
    \label{fig:FIGS1}
\end{figure*}
As shown in Fig.~\ref{fig:FIGS1}, increasing $R$ systematically improves the performance of PA relative to SA (for both cases), progressively reducing the residual energy gap $\Delta E$. To appreciate the improvements in energy, the reader should focus on the heights of the steps, which systematically decrease when $R$ grows. In the three-dimensional case, PA even reaches $\Delta E=0$ already for $R=100$. However, this improvement in the energy comes at a substantial computational cost: compared with CluMP, the corresponding increase in the number of elementary operations is of several orders of magnitude.
This overhead originates from the linear scaling of the computational cost with the population size $n_{\rm elem}^{\rm PA} \sim R\,n_{\rm elem}^{\rm SA}$. 
Therefore, although PA would eventually reach $\Delta E=0$ in the large-$R$ limit, the population sizes required to match or exceed CluMP performance become computationally prohibitive within the range of system sizes considered here.
These results highlight a fundamental difference between the two approaches. PA enhances the exploration of a complex landscape by evolving a huge number of replicas in parallel, and this is very costly.
CluMP, instead, exploits the local structure of correlations through adaptive cluster optimization, and this is a much better approach. As a result, CluMP achieves comparable or superior optimization performance at a significantly lower computational cost.

Finally, although the wall-clock time of PA could be reduced through parallelization on GPU architectures, such considerations lie beyond the scope of the present work, as similar hardware acceleration strategies could be applied to the replica-based version of CluMP (R-CluMP).


\section{APPENDIX B: Benchmarking CluMP on low-connectivity random regular graphs}

\begin{figure*}[t]
    \centering
    \includegraphics[width=0.9\textwidth]{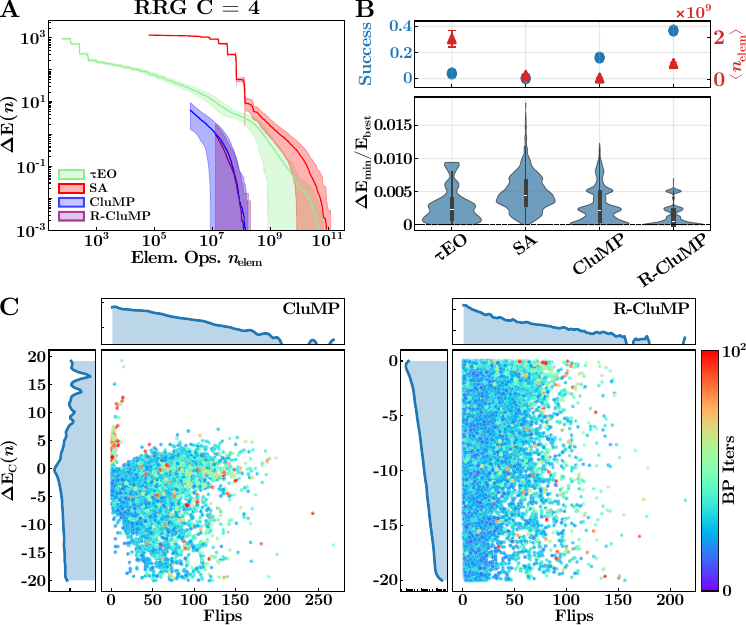}
    \caption{\textbf{CluMP on RRGs with $C=4$ and Gaussian couplings.} Results analogous to Figs. 3 and 4 of the main text. (\textbf{A}) Evolution of the best-so-far energy $\Delta E(n)$ as a function of the number of elementary operations $n_{\rm elem}$. (\textbf{B}) Algorithm performance summarized by the success rate, the average number of elementary operations $\langle n_{\rm elem}\rangle$ required to reach the best observed energy (top), and the distribution of the normalized energy gap $\Delta E_{\min}/E_{\rm best}$ (bottom). (\textbf{C}) Statistics of individual cluster updates for CluMP and R-CluMP, showing the correlation between the cluster energy change $\Delta E_{\mathcal{C}}(n)$ and the number of flipped spins; colors indicate the BP convergence time.  CluMP and R-CluMP yield the best performance among all tested graphs, with the highest success rate ($\sim40\%$) and the fastest convergence ($\sim10^8$ elementary operations).}
    \label{fig:FIGS2}
\end{figure*}

Graph connectivity $C$ plays a crucial role in the performance of the cluster-based optimization strategy. The low-$C$ regime is particularly favorable, as the sparse topology reduces the density of short frustrated loops while preserving a
sufficiently rugged energy landscape. To complement the results at $C=10$ presented in the main text, we benchmark CluMP and R-CluMP against $\tau$ Extremal Optimization ($\tau$EO) and Simulated Annealing (SA) on random regular graphs (RRGs) with connectivity $C=4$ and Gaussian-distributed couplings.
The results, shown in Fig.~\ref{fig:FIGS2}, follow the same analysis as in Figs. 3 and 4 of the main text. CluMP is run for $2^{14}$ iterations, while R-CluMP employs $2^{12}$ iterations with $R=10$ replicas; both algorithms use $\phi=300$. For $\tau$EO, we set $\tau=2.0$, while SA is configured with $(\beta_{\rm final},\, n^{\rm SA})=(10,\, 2^{18})$.

The evolution of the best-so-far energy gap $\Delta E(n)$ as a function of the number of elementary operations $n_{\rm elem}$ (Fig.~\ref{fig:FIGS2}A) shows that both CluMP and R-CluMP significantly outperform the reference single-spin algorithms, displaying a faster relaxation toward the ground state. 
The comparison shown in Fig.~\ref{fig:FIGS2}B summarizes the algorithm performance in terms of success probability, average number of elementary operations $\langle n_{\rm elem}\rangle$ required to reach the best observed energy and distribution of the normalized residual gap $\Delta E_{\min}/E_{\rm best}$.
In this low-connectivity regime, CluMP and R-CluMP achieve their best overall performance across all graph ensembles considered in this work, reaching a success rate of approximately $40\%$ while converging within $\sim10^8$ elementary operations, corresponding to a speedup of $n_{\rm elem} \approx \mathcal{O}(10^4)$ with respect to the competing algorithms.
The statistics of individual cluster updates (Fig.~\ref{fig:FIGS2}C) further clarify the origin of this improved performance. For both CluMP and R-CluMP, large energy decreases are frequently associated with collective rearrangements involving a significant number of spins, indicating that sparse RRGs allow frustration-controlled
clusters to efficiently identify large-scale low-energy directions in
configuration space. Although qualitatively similar behavior is observed at $C=10$, the same effect here occurs at a markedly lower $\phi$, reflecting the larger accessible cluster size before the frustration threshold is reached, ultimately allowing the cluster dynamics to explore broader regions of configuration space through coherent spin rearrangements.


\section{APPENDIX C: Collective Moves on Graphs with Integer Couplings}

In contrast to graphs with Gaussian-distributed couplings, the advantage of CluMP is significantly reduced in the presence of discrete interactions. This behavior reflects a qualitative modification of the underlying energy landscape induced by integer couplings. In fact, for $J_{ij}\in\{\pm1\}$ the local fields
\begin{equation}
h_i = \sum_{j\in\partial i} J_{ij} s_j
\end{equation}
are integer-valued, so exact cancellations ($h_i=0$) occur with finite probability. An extensive fraction of spins is therefore marginal, and both energies and energy differences are discrete. Many configurations are exactly degenerate or separated by the minimal energy gap, generating broad plateaus in the configuration space \cite{mezard1987spin,hartmann1999ground}. Single-spin dynamics naturally exploits these flat directions: zero-field spins can flip at no cost, thereby enabling efficient diffusion across degenerate regions and a gradual local reorganization of frustration. 

In Supplementary Fig.~\ref{fig:FIGS3}, we show the performance of CluMP on random regular graphs (RRGs) with $C=4$ and $C=10$ and $J_{ij}\in\{\pm1\}$, compared to single-spin flip algorithms ($\tau$EO and SA). 
By construction, CluMP performs correlated flips of extended spin sets. In a landscape dominated by discrete cancellations, collective moves frequently involve multiple marginal variables and yield compensating energy contributions, resulting in small gains or exact ties. The resulting reduction in effective selection pressure weakens directed descent. 
This effect becomes even more pronounced at $C=10$ (Fig.~\ref{fig:FIGS3}B). Although the graph connectivity is increased, the degeneracies associated with integer couplings persist, giving rise to a large multiplicity of competing near-degenerate states. As a result, the relative improvement of CluMP remains substantially smaller than in the Gaussian case.
\begin{figure*}[t]
    \centering
    \includegraphics[width=0.9\textwidth]{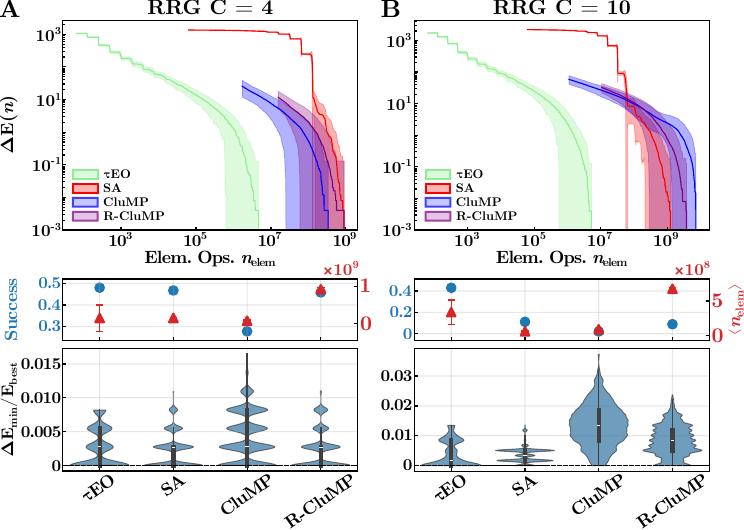}
    \caption{\textbf{CluMP on Random Regular Graphs with Integer Couplings.} Performance of $\tau$EO, SA, CluMP, and R-CluMP on RRGs with integer couplings $J_{ij}\in\{\pm1\}$ and connectivity $C=4$ (\textbf{A}) and $C=10$ (\textbf{B}). Ten disorder samples per graph family, $10^2$ runs per sample. Top: resampled convergence trajectories of the best-so-far energy vs elementary operations $n$ (see main text). Middle: success rate (blue) and average number of effective single-spin iterations to reach $E_{\rm best}$ (red). Bottom: violin plots of the normalized gap $\Delta E_{\min}/E_{\rm best}$ with embedded box plots (median, IQR, $1.5\times$ IQR whiskers). Notice that CluMP and R-CluMP remain competitive with SA, but their advantage over single-spin dynamics is reduced compared to the Gaussian case. The discrete, highly degenerate landscape of integer couplings generates broad plateaus that limit the benefit of correlated cluster updates, particularly at $C=10$.}
    \label{fig:FIGS3}
\end{figure*}

Despite this, CluMP and R-CluMP remain competitive with SA, while they are outperformed by $\tau$EO. Our results indicate that the relative advantage of CluMP depends sensitively on the microscopic structure of the disorder: continuous couplings lift degeneracies and produce a smoother hierarchy of states, whereas integer couplings induce plateau-dominated landscapes that reduce the benefit of correlated cluster updates without eliminating it.


\bibliography{Bibliography}